\documentclass[10pt, twocolumn]{revtex4}
\usepackage{bm,graphicx}
\usepackage[english]{babel}
\usepackage{amssymb, amsthm, amsmath}

\newcommand{\finpr}{\hfill $\square$ \vspace{2mm}}

\def\be{\begin{eqnarray}}

\def\ee{\end{eqnarray}}

\def\bee{\begin{eqnarray*}}

\def\eee{\end{eqnarray*}}

\newtheorem{thm}{Theorem}

\newtheorem{cor}{Corollary}

\newtheorem{lem}{Lemma}

\begin{document}
\title{\bf A monomial matrix formalism to describe quantum many-body states}
\author{Maarten Van den  Nest}
\affiliation{Max-Planck-Institut f\"ur Quantenoptik, Hans-Kopfermann-Str. 1, D-85748 Garching, Germany.}

\begin{abstract}
We propose a framework to describe and simulate a class of many-body quantum states. We do so by considering joint eigenspaces of  sets of monomial unitary matrices, called here ``M-spaces''; a unitary matrix is \emph{monomial} if precisely one entry per row and column is nonzero.   We show that M-spaces encompass various important state families, such as all Pauli stabilizer states and codes, the AKLT model, Kitaev's (abelian and non-abelian) anyon models, group coset states, W states and the locally maximally entanglable states. We furthermore show how basic properties of M-spaces can transparently be understood by manipulating their monomial stabilizer groups. In particular we derive a unified procedure to construct an eigenbasis of any M-space, yielding an explicit formula for each of the eigenstates.  We also discuss the computational complexity of M-spaces and show that basic problems, such as estimating local expectation values, are NP-hard. Finally we prove that a large subclass of M-spaces---containing in particular most of the aforementioned examples---can be simulated efficiently classically with a unified method.
\end{abstract}

\maketitle

\section{Introduction}

The Pauli stabilizer formalism (PSF) is an important tool in quantum information theory. This formalism regards many-body quantum states, called Pauli stabilizer states,  that occur as joint eigenstates of sets of commuting Pauli operators. By exploiting the description of states in terms of their stabilizers, the PSF provides a powerful  method to analyze the properties and dynamics of stabilizer states in a variety of settings---in fact, the PSF is commonly used in virtually all subfields of quantum information. Important applications include quantum error-correction \cite{Go97}, measurement-based computing \cite{Ra01} and classical simulations of quantum  circuits \cite{Go98}. In addition, the PSF is used in condensed matter physics, cf. the study of topological order \cite{Ki03}.

Notwithstanding its success, a drawback of the PSF is that it describes a relatively small class of states. In particular,  there are only finitely many stabilizer states for each given system size. Furthermore these states  have very particular properties. For example, (modulo some trivial cases) they cannot be unique ground states of two-body Hamiltonians,  every qubit is either maximally entangled with the rest of the system or completely disentangled from it, most interesting Pauli stabilizer states have zero correlation length etc. \cite{Va08, He06}. In addition, by definition the PSF only regards commuting stabilizers operators.
This situation prompts the question of whether it is possible to enlarge the class of stabilizer states while maintaining a transparent stabilizer-type description. Such generalizations could lead to new insights in many-body quantum states as well as novel applications, such as better error-correcting codes, new information-theoretic protocols and new quantum states/processes that can be simulated efficiently classically.

A central feature of the PSF is that relevant information about stabilizer states can transparently and efficiently be extracted by suitably manipulating their stabilizer groups. Aiming at generalizing the PSF, our goal is to identify a mathematical structure which is richer than the PSF while maintaining similarly clearcut maps from the ``stabilizer picture'' to the ``state picture''. Our route towards this end starts with the following observation:   all Pauli operators are \emph{monomial} matrices i.e. precisely one matrix entry per row and per column is nonzero.  The basic premise of this work is then to consider  \emph{arbitrary} monomial unitary operators (with efficiently computable matrix elements) as stabilizer operators, giving rise to a general ``monomial stabilizer formalism'' (MSF).

Part of the initial motivation for considering the MSF stems from the fact that monomial matrices are simple operators with favorable mathematical properties. In particular, any group \emph{generated} by a set of monomial matrices consists \emph{entirely} of monomial matrices, making such groups rather manageable objects. Most of the motivation for studying the MSF, however, is an a posteriori one; in fact one of the purposes of the present work is to argue that the MSF has several nice features and to make a case for the further investigation of this framework as an interesting generalization of the PSF. We will do so by means of the following two contributions (cf. section \ref{sect_summary} for a more detailed summary):

\

(a) We show that---perhaps surprisingly---a variety of important quantum many-body states are covered by the MSF, demonstrating that the latter is significantly richer than the PSF.

\

(b) We show that basic properties of M-spaces can be transparently described  by  manipulating their monomial stabilizer groups; this will lead in particular to efficient classical simulations of a subclass of M-states.

\

In view of these features, and taking into account the wide applicability of the PSF, we believe that the MSF provides a promising avenue to describe and simulate interesting many-body states, potentially leading to new applications. Finally we refer to an upcoming work \cite{Va11} where the methods developed in the present work are used to arrive at new efficient classical simulations of quantum Fourier transforms.

\section{Summary of results}\label{sect_summary}

Next we summarize the main results of this work, outlining in particular the content of contributions (a) and (b) mentioned above.

Generalizing the notion of a Pauli stabilizer code, the +1 common eigenspace of a set of monomial unitary operators will be called an M-space. If an M-space is one-dimensional, its (up to a global phase) unique  element will be called an M-state, generalizing the notion of Pauli stabilizer states.

\vspace{2mm}

As for (a) we will demonstrate that the MSF encompasses, in addition to all Pauli stabilizer states and codes, the following important state families: the ground level of the Affleck-Kennedy-Lieb-Tasaki model \cite{Af88}; the ground levels of Kitaev's quantum double models, which describe both Abelian and non-Abelian anyonic systems \cite{Ki03}; the Laughlin wavefunction at filling fraction $\nu=1$ \cite{La83}; the family of locally maximally entanglable (LME) states \cite{Kr08};  coset states of Abelian groups (cf. the Abelian hidden subgroup problem \cite{Jo01}); coherent probabilistic computations \cite{Br09}; W states \cite{Du00}; Dicke states \cite{Di54}.

These examples demonstrate the richness of the MSF. They also show that M-states generally do not display the aforementioned ``special'' features of Pauli stabilizer states. For example there do exist interesting M-states which have non-commuting stabilizer groups and which are unique ground states of two-body Hamiltonians.

\vspace{2mm}

As for (b) we will establish basic maps from the stabilizer picture to the state picture by showing how a designated orthonormal basis (called here the \emph{orbit basis}) of any M-space can be constructed when the latter is described in terms of a set of monomial stabilizers. The procedure yields an explicit formula for each basis state, formulated entirely in terms of manipulations on the stabilizer group.
This result applies to arbitrary M-spaces and thus in particular to all examples given above. In other words one obtains a single unified method to treat a number of seemingly unrelated state families. It will also follow from our analysis that all M-states have a common, particularly simple structure viz. the nonzero-amplitudes of any  M-state are all equal in modulus.

We subsequently use the orbit basis construction to investigate classical simulations. Whereas within the PSF many quantities of interest can be computed efficiently for \emph{all} Pauli stabilizer states, the situation is  different for general M-spaces. We will show that one cannot hope for general efficient algorithms for several basic problems (such as estimating local density operators), as we will prove their NP-hardness. In other words the MSF is too rich a framework to allow for generally applicable efficient simulations. However, it is important that these results regard worst-case complexity. In fact, based on our characterization of the orbit basis, we will identify a relevant subclass of M-states for which \emph{efficient} classical simulations can nonetheless be achieved; this subclass contains in particular almost all examples listed in (a).

Throughout the paper we will illustrate our general constructions with examples. It is noteworthy that our methods allow to recover, with one unified method, the classical simulatability of a variety of state families including the Pauli stabilizer states \cite{Go98} and the quantum double models \cite{Ag08, Vi08}.  In addition, the MSF allows to rederive, in a new and unified way, the standard basis expansion of stabilizer states in terms of cosets of $\mathbb{Z}_2$-linear spaces \cite{De03} as well as the matrix product state basis of the ground level of the AKLT model \cite{Af88}.

\section{Notations and conventions}
All Hilbert spaces considered in this work are finite-dimensional. We will often consider unnormalized quantum states in order not to overload the notation. The group ${\cal G}$ generated by a set of operators $U_1, \dots, U_m$ is denoted by ${\cal G}=\langle U_1, \dots, U_m\rangle$. We also remark that some proofs will be presented in appendices.

\section{M-states and M-spaces}\label{sect_monomial_unitary}

 Let ${\mathcal H}$ be a Hilbert space with orthonormal basis \be{\mathcal B}:=\{|x\rangle: x\in {\mathcal I}\},\ee were ${\mathcal I}$ denotes some finite set. We will mostly consider ${\cal H}$ to be a multi-party tensor product of $d$-dimensional local spaces and ${\cal B}$ will usually be a product basis, but our arguments hold for arbitrary spaces and bases. A unitary operator is called ${\mathcal B}$-monomial if it can be written as a product $U=PD$ where $D$ is a diagonal unitary operator in the basis ${\cal B}$ (thus containing phases on its diagonal) and where $P$ is a permutation matrix in this basis. Equivalently, the matrix representation of $U$ in the basis ${\cal B}$ contains precisely one nonzero entry per row and per column.  If $U$ and $U'$ are ${\cal B}$-monomial operators, so are the operators $U^{\dagger}$ and $UU'$. Furthermore if $U$ is ${\cal B}$-monomial and $V$ is ${\cal B}'$-monomial then $U\otimes V$ is monomial relative to the tensor product basis ${\cal B}\otimes {\cal B'}$.

In the following it will usually be clear from the context which basis ${\cal B}$ is considered. Therefore,  the prefix ``${\cal B}$-'' will mostly be omitted i.e. we will simply refer to ``monomial'' operations. Elements of ${\cal B}$ will typically be denoted by $|x\rangle$, $|y\rangle$, $|z\rangle$.

An important feature of  monomial operations is that  products $U_1U_2\dots U_t$ remain monomial regardless of the length of the product, as long as every $U_i$ is monomial (relative to the same basis). This implies in particular that if a group is \emph{generated} by a set of monomial operators, then \emph{all} elements in this group are monomial.

The +1 common eigenspace of a set $\{U_1\dots, U_m\}$  of operators is the space of all $|\psi\rangle$ satisfying \be\label{M-space} U_i|\psi\rangle=|\psi\rangle\mbox{\quad  for every } i=1, \dots, m.\ee A space ${\cal M}$ is called a monomial unitary stabilizer space, or M-space in short, if there exists a set $\{U_i\}$ of monomial unitary operators with +1 common eigenspace ${\cal M}$. The group  ${\cal G}=\langle U_1, \dots, U_m\rangle$  is said to be  a monomial stabilizer group of ${\cal M}$ \footnote{Remark that distinct monomial stabilizer groups may give rise to the same M-space.}.  A state $|\psi\rangle$ is called a monomial unitary stabilizer state, or simply M-state, if there exists a set of unitary monomial operators which have this state as their unique +1 common eigenvector, up to a global phase. In this paper we will only consider \emph{finite} stabilizer groups ${\cal G}$ \footnote{However, many of the results in this work naturally generalize to infinite stabilizer groups.}. Otherwise, in the above definitions no restrictions are  placed on the operators  $U_i$ except their unitarity and monomiality.

In the following we envisage M-spaces ${\cal M}$ which are defined in terms of a set of equations  (\ref{M-space}). A basic problem will then be to understand how features of ${\cal M}$ can be traced back to features of  ${\cal G}$. More stringently, we will study the \emph{computational hardness} of determining certain quantities (such as expectation values of local observables) by suitably manipulating the generators $U_i$. To make meaningful statements about computational efficiency issues, it needs to be made clear which classical descriptions of the $U_i$ are considered to be available (in particular one should restrict to M-spaces which have \emph{efficient} descriptions), and how computational cost is measured. This discussion is postponed to section \ref{sect_complexity}. In sections \ref{sect_examples}-\ref{sect_main_results} we will not yet worry about efficiency issues and our treatment will hold for M-spaces in general.

\section{Examples}\label{sect_examples}

The paradigm of M-states and -spaces encompasses a number of important quantum many-body states:

\begin{itemize}

\item Pauli stabilizer states \cite{He06} and codes \cite{Go97}. Important examples include the cluster states \cite{Br01}, the Greenberger-Horne-Zeilinger states \cite{Gr89} and the toric code \cite{Ki03}.

\item Generalized Pauli stabilizer states and codes for $d$-level systems \cite{Go99}.

\item The Affleck-Kennedy-Lieb-Tasaki model \cite{Af88}.

\item Kitaev's quantum double models. These are generalizations of the toric code which describe systems of abelian and non-abelian anyons \cite{Ki03}.

\item Laughlin's wavefunction at filling fraction $\nu=1$ \cite{La83}.

\item Locally maximally entanglable (LME) states \cite{Kr08}. This class was recently
 introduced in studies of multipartite entanglement.

\item Coherent probabilistic computations \cite{Br09}. These states represent the natural embedding of  probabilistic classical computation (i.e. BPP) into quantum computation (BQP).

\item Coset states of finite Abelian groups. These states occur in important quantum algorithms viz. the Abelian hidden subgroup problem (see e.g. \cite{Jo01}).

\item W-states \cite{Du00} and more generally Dicke states \cite{Di54}.

\end{itemize}
Remark that the above examples occur in different areas of application, ranging from topologically ordered systems (cf. quantum double models) to quantum algorithms (cf. coset states) to multipartite entanglement studies (cf. W states and LME states). This richness of M-spaces motivates their study as a general framework, as initiated in the present paper. We briefly discuss the Pauli stabilizer states/codes, the AKLT model and the LME states here. See  appendix \ref{app_sect_further_examples} for a discussion of the other examples.

\subsection{Pauli stabilizer states and codes}

An $n$-qubit  Pauli operator has the form $\sigma= \gamma \sigma_1\otimes\dots\otimes \sigma_n$ where each  $\sigma_k$ is either the single-qubit identity operator or one of the Pauli matrices and where $\gamma= \pm 1, \pm i$. A linear subspace of a $n$-qubit system is a Pauli stabilizer code if there exists a set of commuting Pauli operators that has this space as its $+1$ common eigenspace. If a Pauli stabilizer code is one-dimensional then its unique (up to a global phase) element is called a Pauli stabilizer state. Remark that the Pauli matrices and the identity are obviously monomial (relative to the standard basis). Since tensor products of monomial matrices are again monomial, we find that every Pauli operator is unitary and monomial relative to the computational basis. Thus every Pauli stabilizer code/state is an M-space/state.

Remark that general monomial unitary stabilizer groups are significantly richer than the Pauli stabilizer groups. For example, Pauli operators are \emph{product operators}, whereas elements of general monomial stabilizer groups need not be. Second, Pauli stabilizer groups are \emph{Abelian}, whereas general monomial stabilizer groups can be non-Abelian (cf. the AKLT model in section \ref{sect_AKLT}). Finally, the number of Pauli stabilizer groups (and thus also the number of Pauli stabilizer states/codes) is \emph{finite} for a given Hilbert space dimension. Monomial unitary stabilizer groups on the other hand form a continuous family.

Owing to the richer structure of the allowed stabilizer groups, M-states/spaces may exhibit features that can never be present in Pauli stabilizer states/codes. For example, for every Pauli stabilizer state the bipartite entanglement between any qubit and the rest of the system is either maximal or zero \cite{He06}.  Also the localizable entanglement \cite{Ve04} between any two qubits is either zero or maximal \cite{He06}. It is however easy to give examples of M-states where these entanglement measures vary continuously. It is also known that, except for some uninteresting cases, no Pauli stabilizer state can be the unique ground state of a two-body Hamiltonian \cite{Va08}. The example of the AKLT model (cf. section \ref{sect_AKLT}) shows however that there do exist interesting M-states with this property.

\subsection{AKLT model}\label{sect_AKLT}

We show that the ground level of the one-dimensional spin-$1$ AKLT model is an M-space; to our knowledge this is a new way of describing the AKLT model.

Consider two spin-$1$ particles with local basis $\{|0\rangle, |1\rangle, |2\rangle\}$. Let $\pi$ be the  projector onto the subspace spanned by \be\label{AKLT_psi_i} \begin{array}{ccl} |\psi_1\rangle =|01\rangle- |10\rangle &\quad& |\psi_3\rangle =|12\rangle- |21\rangle\\ |\psi_2\rangle = |02\rangle- |20\rangle && |\psi_4\rangle =|00\rangle+ |11\rangle + |22\rangle. \end{array}\ee
Now consider a  1D chain of spin-$1$ particles labelled from 1 to $n$ where one may consider either periodic or open boundary conditions.  Let $H_{i,i+1}$ act as $I-\pi$ on spins $i$ and $i+1$ and remark that $H_{i,i+1}\geq 0$. Let ${\cal M}$ denote the zero energy ground state subspace of the AKLT Hamiltonian $H_{\mbox{\scriptsize{aklt}}}=\sum H_{i,i+1}$ \cite{Af88}. Since $H_{i,i+1}\geq 0$, the space ${\cal M}$ coincides with the the 0 common eigenspace of the operators $H_{i,i+1}$. In the case of open boundary conditions,  ${\cal M}$ is four-fold degenerate; in the case of periodic boundary conditions, ${\cal M}$ is one-dimensional i.e. the AKLT ground state is unique \cite{Af88}.

Let $U$ be the following monomial unitary operator: \be\label{AKLT_U} U\ :\ \begin{array}{lcl} |01\rangle \leftrightarrow -|10\rangle&\quad &|12\rangle\leftrightarrow -|21\rangle\\ |02\rangle\leftrightarrow -|20\rangle &\quad &|00\rangle\to |11\rangle\to |22\rangle \to |00\rangle.\end{array}\ee
It is straightforward to show that the $+1$ eigenspaces of $U$ and $\pi$ coincide. Letting $U_{i,i+1}$ act as $U$ on spins $i$ and $i+1$ it follows that ${\cal M}$ is the +1 common eigenspace of the $U_{i,i+1}$ and thus an M-space.

\subsection{LME states}

An $n$-qubit state $|\psi\rangle$ is  LME (locally maximally entanglable) if there exists a unitary product operator $U=U_1\otimes\dots \otimes U_n$ such that \be\label{LME} U|\psi\rangle\propto \sum\gamma_x |x\rangle\ee for some complex phases $\gamma_x$, where the $|x\rangle$ denote $n$-qubit computational basis vectors and where the sum is over the entire basis. LME states were recently introduced in \cite{Kr08} where it was shown that $|\psi\rangle$ is LME if and only if this state can be maximally entangled to an ancillary $n$-qubit system using local controlled-unitary operations \footnote{The latter is actually the original definition of LME states.}. It was also shown that the family of LME states contains all Pauli stabilizer states. In turn, all LME states are M-states. This readily follows from the stabilizer description of LMEs introduced in \cite{Kr08}, which we briefly repeat here. Define the diagonal unitary operator $D:= \sum \gamma_x|x\rangle\langle x|$ and let $X_i$ be be the Pauli $X$ operator acting on qubit $i$. Since $|+\rangle: = |0\rangle +|1\rangle$ is the unique +1 common eigenvector of the $n$ operators $X_i$, it follows that that $|\psi'\rangle := D|+\rangle^{n} = \sum \gamma_x |x\rangle$ is the unique +1 common eigenvector of the operators $U_i:= DX_iD^{\dagger}$. Remark that the $U_i$ are unitary and monomial relative to the computational basis so that $|\psi'\rangle$ is an M-state. As $|\psi\rangle = U|\psi'\rangle$ this immediately implies that $|\psi\rangle$ is an M-state relative to the product basis ${\cal B}:=\{U|x\rangle\}$.

\section{Preliminary concepts}\label{sect_basic}

Here we introduce some concepts that will be needed in the statement and proofs of our main results below.

\subsection{The projector $\rho$}\label{sect_basic_rho}

Consider a finite unitary  group ${\cal G}$ with +1 common eigenspace ${\cal M}$ and let $\rho$ denote the orthogonal projector onto ${\cal M}$. Then $\rho$ coincides with the ``group averaging operator'' \be\label{rho} \rho = \frac{1}{|{\cal G}|} \sum_{U\in{\cal G}} U.\ee  Furthermore one has \be\label{U_rho=rho} U\rho = \rho = \rho U\mbox{\quad for every } U\in {\cal G}.\ee
For completeness, proofs of these properties are given in appendix \ref{app_sect_rho}.

\subsection{The support of a subspace} Consider a Hilbert space ${\cal H}$ with orthonormal basis ${\cal B}$. The  ${\cal B}$-support, or simply the support,  of a state $|\psi\rangle$ is the set of basis states $|x\rangle\in{\cal B}$ for which $\langle x|\psi\rangle$ is nonzero. The support of a linear subspace ${\cal M}$ of ${\cal H}$ is the union of the supports of its elements. The support of a state/subspace will be denoted by supp$(|\psi)$) and supp$({\cal M})$, respectively.

Letting $\rho$ denote the orthogonal projector on the space ${\cal M}$, it is easily verified that the following statements are equivalent:

\vspace{2mm}

(a) $|x\rangle\in $ supp$({\cal M})$; \  (b) $\rho|x\rangle\neq 0$; \ (c) $\langle x|\rho|x\rangle\neq 0$.

\subsection{Uniform superpositions}\label{sect_uniform_sup}

Consider a subset  ${\mathcal O}\subseteq {\mathcal B}$ and  a  complex phase $\gamma_y$ for every $|y\rangle\in{\cal O}$. Consider the states \be\label{uniform_sup} |\psi\rangle = \frac{1}{\sqrt{|{\cal O}|}}\sum_{|y\rangle\in {\cal O}} \gamma_y |y\rangle, \quad |{\cal O}\rangle = \frac{1}{\sqrt{|{\cal O}|}}\sum_{|y\rangle\in {\cal O}} |y\rangle.\ee Any state  as in the l.h.s. of (\ref{uniform_sup}) is called a ${\cal B}$-uniform superposition, or simply a uniform superposition.  Remark that  ${\cal O}$ is precisely the support of $|\psi\rangle$. The state $|{\cal O}\rangle$ is called the equal superposition over ${\cal O}$.

\subsection{The permutation group $ {\cal P}$}\label{sect_bar_U_bar_G}

Consider a unitary monomial operator $U=PD$ where $P$ is a permutation and $D$ is diagonal; note that this decomposition is unique. Henceforth we will denote $\bar U:=P$. Let ${\cal G}=\langle U_1, \dots, U_m\rangle$ be a  unitary monomial group. The set  ${\cal P}= \{ \bar U: U\in {\cal G}\}$ is called the permutation group of ${\cal G}$. Using that $ \overline{ UV}= \bar U\bar V$ for every unitary monomial $U$ and $V$, it can easily be shown that ${\cal P}$ is a group generated by the operators $\bar U_i$ \footnote{In fact the map $\varphi: U\to \bar U$ is a group homomorphism.}.
The group ${\cal P}$ naturally acts as a permutation group on ${\cal B}$. Consider the  orbit of $|x\rangle$: \be {\cal O}_x = \{|y\rangle: \exists P\in{\cal P}\mbox{ s.t. } P|x\rangle=|y\rangle\}.\ee   It is well-known that every $|x\rangle$ belongs to precisely one orbit (i.e. the orbit ${\cal O}_x$).
Straightforwardly applying the definition of ${\cal P}$ one finds that $|y\rangle\in {\cal O}_x$ if and only if there exists a complex phase $\xi$ and  $U\in{\cal G}$ such that $U|x\rangle  = \xi|y\rangle$. This elementary property will often be used in the following.

\

{\it Example.} Throughout this section as well as section \ref{sect_main_results} we illustrate the various concepts with a simple ``running example''. Consider the $n$-qubit diagonal operator \be T := \bar\alpha\cdot \Lambda^{\otimes n}; \quad \Lambda := \mbox{ diag}(1, \alpha), \quad \alpha := e^{\frac{2\pi i}{n}}.\ee Further, let $S_i$ denote the  SWAP gate acting on qubits $i$ and $i+1$, where $i=1, \dots, n-1$. The operators $T$ and $S_i$ are unitary and monomial (relative to the computational basis). We consider the group ${\cal G}_w$ generated by these operators. Since $T$ is diagonal, one has $\bar T=I$. Since $S_i$ is a permutation matrix, one has $\bar S_i = S_i$. Thus ${\cal P}_w$ is generated by the operators $S_i$. It is easily verified that ${\cal P}_w$ has $n+1$ orbits ${\cal O}_0, \dots, {\cal O}_{n}$, where ${\cal O}_i$ contains all computational basis states $|x\rangle$ where $x$ is a bit string with Hamming weight $i$ (that is, precisely $i$ entries are equal to 1). \hfill $\diamond$

\subsection{The phases $\xi_x(y)$}\label{sect_basic_xi}

Let ${\cal M}$ be an M-space with stabilizer group ${\cal G}$ and let $\rho$ be the orthogonal projector on ${\cal M}$. Fix $|x\rangle\in$ supp$({\cal M})$ (hence $\rho|x\rangle$ is nonzero) and $|y\rangle\in {\cal O}_x$ arbitrarily. Then there exists $U\in{\cal G}$ and a complex phase $\xi$ such that $U|x\rangle = \xi |y\rangle$.  We consider the set $\Xi_{x\to y}$ of all phases that may occur in this way: \be\label{set_xi} \Xi_{x\to y}:=\{\xi: \exists U\in{\cal G} \mbox{ s.t. } U|x\rangle = \xi|y\rangle\}.\ee Remarkably, this set is in fact a \emph{singleton}. To see this, consider $\xi \in \Xi_{x\to y}$ and a corresponding $U\in {\cal G}$. Using that $\rho = \rho U$ it follows that $\rho |x\rangle = \xi\rho |y\rangle.$ Thus $\rho |y\rangle$ is proportional to $\rho |x\rangle$ and the proportionality factor is precisely given by $\xi$. This shows that this phase does not depend on $U$, so that $\Xi_{x\to y}$ is indeed a singleton. In the following we will denote the unique element of this set by $\xi_x(y)$. We have proved:
\begin{lem}\label{thm_xi}
For every $|x\rangle\in$ supp$({\cal M})$ and $|y\rangle\in {\cal O}_x$ one has $\rho|x\rangle = \xi_x(y)\rho|y\rangle$.
\end{lem}
We emphasize that the phase $\xi_x(y)$ is only defined for $|x\rangle\in$ supp(${\cal M}$) and $|y\rangle\in {\cal O}_x$.  Finally, note that the phase $\xi_x(x)$ is well-defined for every $|x\rangle$ in the support of ${\cal M}$, since $|x\rangle\in {\cal O}_x$. Using that $I\in {\cal G}$ and $I|x\rangle = |x\rangle$, we in fact find that $\xi_x(x)=1$.

\

{\it Example.} Consider the group ${\cal G}_w$ as above and let ${\cal M}_w$ be its +1 common eigenspace. We show in section \ref{sect_main_results} that the basis state $|e_1\rangle:=|10\cdots 0\rangle$ belongs to the support of ${\cal M}_w$. Note that $|e_1\rangle$ belongs to the orbit ${\cal O}_1$. The other vectors in this orbit are $|e_2\rangle, \dots, |e_n\rangle$ where $e_i $ denotes an $n$-bit string with a 1 in the $i$-th slot and zeroes elsewhere. Letting $P_i$ denote the operator which swaps qubits $1$ and $i$ (which can easily be obtained as a suitable product of $S_k$ gates) one has $P^i|e_1\rangle= |e_i\rangle$. It follows that $\xi_{e_1}(e_i)=1$ for every $i$.\hfill $\diamond$

\section{The orbit basis}\label{sect_main_results}

Consider an M-space ${\cal M}$ specified in terms of a stabilizer group ${\cal G}$. The latter may e.g. be given in terms of a set of generators. Our goal is to construct an orthonormal basis of ${\cal M}$ assuming no prior information about this space except for the group ${\cal G}$.
In this section we prove two results which will directly lead to such a construction.
The first of these theorems characterizes the support of ${\cal M}$, the second theorem  characterizes a designated basis of ${\cal M}$ called the \emph{orbit basis}.

\subsection{Characterizing the support}\label{sect_support}

We need some  notation. For every $|x\rangle\in {\cal B}$ let ${\cal G}_x$ be the set of all $U\in{\cal G}$ which have $|x\rangle$ as an eigenvector. This set is a subgroup of ${\cal G}$. We let $\{U_{x, 1}, \dots, U_{x, l}\}$ be an arbitrary set of generators \footnote{The number of generators $l$ generally depends on $x$ but this dependence is suppressed not to overload notation.} of ${\cal G}_x$.

\begin{thm}[\bf{Support of M-space}]\label{thm_support}
Consider an M-space ${\cal M}$ with stabilizer group ${\cal G}$. There exist orbits ${\cal O}_i$ such that $\mbox{supp}({\cal M})= {\cal O}_{1}\cup\cdots\cup {\cal O}_{d}$. Furthermore consider an arbitrary $|x\rangle\in {\cal B}$. Then the following statements are equivalent:

\begin{itemize}

\item[(a)] ${\cal O}_x\subseteq$ supp(${\cal M}$).

\item[(b)] $\langle x|U|x\rangle\in\{0, 1\}$ for every $U\in {\cal G}$.

\item[(c)] $U|x\rangle = |x\rangle$ for every $U\in{\cal G}_x$.

\item[(d)] $U_{x, i}|x\rangle = |x\rangle$ for every $i$.
\end{itemize}
\end{thm}
{\it Proof: } lemma \ref{thm_xi} shows that for every $|x\rangle$ in the support of ${\cal M}$ and $|y\rangle\in {\cal O}_x$, one has $\rho|y\rangle\neq 0$ so that also $|y\rangle\in$ supp$({\cal M})$. This shows that the entire orbit ${\cal O}_x$ must be contained in the support. Thus supp$({\cal M})$ is a union of orbits as claimed.

{\bf [a}$\mathbf{\Rightarrow}${\bf c]} If ${\cal O}_x\subseteq$ supp(${\cal M}$) then $|x\rangle$ belongs to the support of ${\cal M}$ since $|x\rangle\in {\cal O}_x$. For every $|x\rangle$ in the support the phase $\xi_x(x)$ is well-defined and in fact $\xi_x(x)=1$ (cf. section \ref{sect_basic_xi}).  Consider an arbitrary $U\in{\cal G}_x$ i.e. there exists a phase $\xi$ such that $U|x\rangle = \xi|x\rangle$. By uniqueness of the phase $\xi_x(x)$ it follows that $\xi= \xi_x(x)=1$.

 {\bf [c}$\mathbf{\Rightarrow}${\bf b]}
Consider an arbitrary $U\in {\cal G}$. Then $U|x\rangle = \xi|y\rangle$ for some complex phase $\xi$ and some $|y\rangle$. If $y\neq x$ then $\langle x|U|x\rangle $ is zero. If $y=x$ then $U\in {\cal G}_x$ and thus $\xi$ equals +1 by assumption (c). Thus $\langle x|U|x\rangle = \xi = 1$.

{\bf [b}$\mathbf{\Rightarrow}${\bf a]} Let $\rho$ denote the orthogonal projector onto ${\cal M}$. Using (\ref{rho}) we have \be \langle x|\rho|x\rangle = \frac{1}{|{\cal G}|}\sum \langle x|U|x\rangle. \ee
Owing to (b) every term  $\langle x|U|x\rangle$ in the sum is nonnegative. Moreover there is at least one nonzero term i.e. when $U$ is the identity. This shows that $\langle x|\rho|x\rangle$ is nonzero so that $|x\rangle$ belongs to the support of ${\cal M}$. But then ${\cal O}_x\subseteq$ supp(${\cal M}$) since the support is a union of orbits.

{\bf [c}$\mathbf{\Leftrightarrow}${\bf d]} This equivalence is straightforward since the $U_{x, i}$ generate ${\cal G}_x$.
\finpr

{\it Example. } Consider the M-space ${\cal M}_w$ with stabilizer group ${\cal G}_w$ as in section \ref{sect_basic}. For every computational basis state $|x\rangle$ one has \be T|x\rangle = \bar \alpha\cdot  \alpha^{|x|}|x\rangle\ee where $|x|$ denotes the Hamming weight of $x$. This shows that, for every $x$ with $|x|\neq 1$, one has $T|x\rangle = \lambda|x\rangle$ for some $\lambda\neq 1$. Invoking theorem \ref{thm_support}, we find that none of the orbits ${\cal O}_i$ with $i\neq 1$ are contained in the support of ${\cal M}_w$. On the other hand, the orbit ${\cal O}_1$ is contained in the support. To show this, consider the $n$-qubit W state \be |W\rangle:=\frac{1}{\sqrt{n}}[|e_1\rangle + \cdots + |e_n\rangle]\ee It is straightforward to verify that $T|W\rangle = |W\rangle = S_i|W\rangle$ for every $i$, showing that $|W\rangle\in {\cal M}_w$. Since this state has support ${\cal O}_1$, it follows that ${\cal O}_1\subseteq $ supp$({\cal M})_w$.
In conclusion, the support of ${\cal M}_w$ is identical to ${\cal O}_1$.\hfill $\diamond$

\subsection{Constructing the orbit basis}

For every $|x\rangle\in$ supp$({\cal M})$ consider its normalized projection onto ${\cal M}$: \be\label{psi_x} |\psi_x\rangle := \frac{\rho|x\rangle}{\| \rho|x\rangle\|}.\ee Henceforth whenever considering a state (\ref{psi_x}) we will tacitly assume that $|x\rangle$ belongs to the support since otherwise $\rho|x\rangle=0$. Note that by construction $U|\psi_x\rangle = |\psi_x\rangle$ for every $U\in{\cal G}$. 
Interestingly, the states $|\psi_x\rangle$ have the following explicit form:

\begin{lem}[{\bf Orbit states}]\label{thm_psi_x}
Every $|\psi_x\rangle$ is a uniform superposition state given by: \be \label{psi_x_uniform_sup} |\psi_x\rangle = \frac{1}{|{\cal O}_x|^{\frac{1}{2}}}\sum_{|y\rangle\in {\cal O}_x} \xi_x(y) |y\rangle.\ee
\end{lem}
{\it Proof:} using that $\rho$ is given by (\ref{rho}) one has \be\label{psi_x_support}|\psi_x\rangle \propto \sum_{U\in{\cal G}} U|x\rangle =\sum_{|y\rangle\in {\cal O}_x} c_y |y\rangle,\ee for some coefficients $c_y$. This shows that supp$(|\psi_x\rangle)\subseteq {\cal O}_x$. Furthermore for every $|y\rangle\in {\cal O}_x$ there exists $U\in {\cal G}$ such that $U|x\rangle =\xi_x(y)|y\rangle$. Using that $U^{\dagger}|\psi_x\rangle =|\psi_x\rangle$ it follows that \be \langle y|\psi_x\rangle = \xi_x(y) \cdot \langle x|\psi_x\rangle.\ee In combination with (\ref{psi_x_support}) this implies that \be |\psi_x\rangle = \langle x|\psi_x\rangle \sum_{|y\rangle\in {\cal O}_x} \xi_x(y) |y\rangle.\ee Since $|\psi_x\rangle$ is normalized, it follows that \be \langle x|\psi_x\rangle= \frac{\alpha}{\sqrt{|{\cal O}_x|}}\ee for some complex phase $\alpha$. Finally, it follows from definition (\ref{psi_x}) that $\langle x|\psi_x\rangle = \|\rho|x\rangle\|>0$. This shows that $\alpha=1$. \finpr

Since $|\psi_x\rangle$ is a uniform superposition over an orbit of ${\cal P}$, we call this state an \emph{orbit state}. More precisely we  call $|\psi_x\rangle$ the orbit state determined by $|x\rangle$.

\

{\it Example.}  The orbit state of ${\cal M}_w$ determined by $|e_1\rangle$ is  the $n$-qubit W-state:\be |\psi_{e_1}\rangle &=& \frac{1}{\sqrt{|{\cal O}_1|}}\sum_{|y\rangle\in {\cal O}_1} \xi_{e_1}(y)|y\rangle \nonumber \\ &=& \frac{1}{\sqrt{n}}[|e_1\rangle + \cdots + |e_n\rangle]=|W\rangle.\ee
\hfill $\diamond$

\

Next we show that a basis of ${\cal M}$ can be constructed by selecting a suitable subset of orbit states; this is the orbit basis.

\begin{thm}[{\bf Orbit basis}]\label{thm_stab_space}
Consider an M-space ${\cal M}$ with stabilizer group ${\cal G}$ and support $\mbox{supp}({\cal M})= {\cal O}_{1}\cup\cdots\cup {\cal O}_{d}$.  Choose an arbitrary representative $|x_i\rangle$ in each orbit ${\cal O}_i$ and the associated orbit state $|\psi_i\rangle := |\psi_{x_i}\rangle$ for every $i$ from 1 to $d$. Then the set $\Psi:=\{|\psi_{1}\rangle, \cdots, |\psi_d\rangle\}$ is an orthonormal basis of ${\cal M}$.  Furthermore this basis is independent (up to global phases) of the choice of orbit representatives $|x_i\rangle$. We call $\Psi$ the orbit basis of ${\cal M}$.

\end{thm}
{\it Proof:} since ${\cal B}$ is a basis of the Hilbert space and since $|\psi_x\rangle$ is defined as the projection of $|x\rangle$ onto ${\cal M}$, the collection of all orbit states span ${\cal M}$ (although generally these states are not linearly independent). Consider an arbitrary $|x\rangle$ in the support of ${\cal M}$. By construction there exists an $i$ between 1 and $d$ such that $|x\rangle\in {\cal O}_i$. Lemma \ref{thm_xi} shows that $\rho |x\rangle \propto \rho |x_i\rangle$ so that $|\psi_{x}\rangle\propto |\psi_{i}\rangle$. Since the orbit states span ${\cal M}$, it follows that the states $|\psi_i\rangle$ span ${\cal M}$ as well. Since $|\psi_i\rangle$ has the orbit ${\cal O}_i$ as its support owing to (\ref{psi_x_uniform_sup}) and since these orbits are mutually disjoint, the states $|\psi_i\rangle$ are mutually orthogonal. This shows that $\Psi$ is an orthonormal basis. Finally, lemma \ref{thm_xi} shows that the basis $\Psi$ is independent (up to global phases) of the choice of orbit representatives $|x_i\rangle$.
\finpr

Theorems \ref{thm_support} and \ref{thm_stab_space} show that the following procedure allows to correctly identify the orbit basis of ${\cal M}$.

\begin{itemize}
\item Determine all orbits of ${\cal P}$ and consider a representative $x_k$ in each orbit ${\cal O}_k$.

\item For each $k$, decide whether ${\cal O}_k\subseteq\mbox{ supp}({\cal M})$ by means of the characterization in theorem \ref{thm_support}.
\item The orbit basis $\Psi$ is the collection of all orbit states $|\psi_{x_k}\rangle$ for which ${\cal O}_k\subseteq\mbox{ supp}({\cal M})$.
\end{itemize}
As desired, this procedure can be implemented by means of manipulations on the stabilizer group ${\cal G}$. Furthermore each orbit state is itself characterized completely in terms of properties of ${\cal G}$ owing to lemma \ref{thm_psi_x}.

\

{\it Example. } We have seen that supp(${\cal M}_w) = {\cal O}_1$. Theorem \ref{thm_stab_space} thus shows that ${\cal M}_w$ is one-dimensional, with $|\psi_{e_1}\rangle$ as its unique element. Furthermore we have shown that   $|\psi_{e_1}\rangle=|W\rangle$. Thus, the W state is an M-state with stabilizer group ${\cal G}_w$.\hfill $\diamond$

\

Remark that the  orbit basis construction  applies to arbitrary M-spaces and thus in particular to all instances given in section \ref{sect_examples}. This yields one unified method to analyze all these state families; see section \ref{sect_illustrations} for further illustrations.

\subsection{Some corollaries}

Next we discuss some immediate corollaries of theorems \ref{thm_support} and \ref{thm_stab_space}. The first corollary is interesting in that bounds the dimension of ${\cal M}$ by means of a purely combinatorial quantity viz. the number of orbits. Remark that there exist well-developed tools to count/estimate orbits of permutation groups, which may thus be imported into the study of M-spaces.
\begin{cor}[{\bf Dimension}]\label{thm_dimension}
The number of orbits $d$ contained in the  support of ${\cal M}$ (cf. theorem \ref{thm_support}) coincides with the dimension of ${\cal M}$. It follows that this dimension is upper bounded by the total number of orbits of ${\cal P}$.
\end{cor}
The above corollary will be used in section \ref{sect_illustrations} to give a simple proof that the AKLT ground level in the open boundary conditions case is four-fold degenerate.

Second, theorem \ref{thm_stab_space} leads to the following characterization of M-states.

\begin{cor}[{\bf M-states}]\label{thm_stab_state} Every M-state $|\psi\rangle$ is a uniform superposition. More precisely, if $|x\rangle$ is an arbitrary basis state satisfying $\langle x|\psi\rangle\neq 0$, then $|\psi\rangle \propto |\psi_x\rangle$ where $|\psi_x\rangle$ is given explicitly in lemma \ref{thm_psi_x}.
\end{cor}
Remark that the support of any M-state coincides with precisely \emph{one} orbit. This implies in particular that knowledge of a \emph{single} $|x\rangle$ satisfying $\langle x|\psi\rangle\neq 0$ implies complete knowledge of the entire support of $|\psi\rangle$, which then must coincide with ${\cal O}_x$. Theorem \ref{thm_support} may be used to determine a suitable $x$ such that $|\psi\rangle\propto |\psi_x\rangle$.

It is interesting to compare corollary \ref{thm_stab_state} with a characterization of Pauli stabilizer states obtained  in \cite{De03}. Consider a Pauli stabilizer state $|\psi\rangle$ on $n$ qubits. Then there exists a linear subspace $S$ of $\mathbb{Z}_2^n$, an $x\in \mathbb{Z}_2^n$ and complex phases $\xi_y$ such that \be\label{stab_state_uniform_sup} |\psi\rangle = \frac{1}{\sqrt{|S|}} \sum_{y\in S+ x} \xi_y |y\rangle.\ee Thus every Pauli stabilizer states is a uniform superposition, the support of which is identified with a coset $x+S$. Corollary \ref{thm_stab_state} shows that, in fact, all M-states have a basis expansion with an analogous structure. Corollary \ref{thm_stab_state} can in fact be used to rederive (\ref{stab_state_uniform_sup}) in a simple way; see section \ref{sect_illustrations}.

Restricting attention to many-qubit systems there is an noteworthy connection to LME states. In the terminology of the present work, an $n$-qubit state $|\psi\rangle$ is LME iff there exists a product basis ${\cal B}$ such that this state is a ${\cal B}$-uniform superposition where supp$(|\psi\rangle)$ is the \emph{entire} basis ${\cal B}$. Corollary \ref{thm_stab_state} shows that $n$-qubit M-states (when monomiality is considered relative to product bases) can be regarded as generalizations of LME states where uniform superpositions with arbitrary supports are considered.

Finally, we discuss a subclass of stabilizer groups for which the results above can be simplified. A monomial unitary group ${\cal G}$ is called \emph{pure} if it has a generating set of the form $\{P_1, \dots, P_k,$  $\Lambda_1, \dots, \Lambda_m\}$ where every $P_i$ is a permutation  and every $\Lambda_j$ is diagonal. Such a generating set is also called pure.  For every $|x\rangle$ let $|{\cal O}_x\rangle$ denote the equal superposition over the orbit ${\cal O}_x$ (recall  section \ref{sect_uniform_sup}).
\begin{cor}[{\bf Pure stabilizer groups}]\label{thm_pure}
Let ${\cal M}$ be an M-space with pure stabilizer group ${\cal G}$.  Then the orbit basis of ${\cal M}$ has the form $\Psi=\{|{\cal O}_{x_1}\rangle, \dots, |{\cal O}_{x_d}\rangle\}$.
\end{cor}
{\it Proof:} let ${\cal P}$ be the permutation group of ${\cal G}$.  Consider a pure generating set $\{P_i, \Lambda_j\}$ of ${\cal G}$. Since $\bar P_i = P_i$ and  $\bar \Lambda_j = I$, it follows that ${\cal P}$ is generated by the operators $P_i$. This implies that ${\cal P}\subseteq {\cal G}$.  Let $|x\rangle$ belong to the support of ${\cal M}$  and consider an arbitrary $|y\rangle\in {\cal O}_x$. Then there exists $P\in{\cal P}$ such that $P|x\rangle = |y\rangle$. Since $P\in{\cal G}$ and owing to the uniqueness of the coefficient $\xi_x(y) $  it follows that $\xi_x(y) = 1$. Invoking lemma \ref{thm_psi_x} it follows that $|\psi_x\rangle =|{\cal O}_x\rangle$. \finpr

Examples of pure stabilizer groups are numerous: consider e.g. the toric code states, quantum double models, W-states, coherent probabilistic computations and coset states of Abelian groups (cf. section \ref{sect_examples} and appendix \ref{app_sect_further_examples}). We will consider the example of quantum doubles in more detail in section \ref{sect_illustrations}.

\section{Computational complexity and classical simulations}\label{sect_complexity}

Thus far, we have studied general mathematical features of M-spaces, not worrying about the computational complexity of the concepts involved. Here we discuss such issues. For concreteness, we consider many-qubit systems where monomiality is defined relative to the computational basis $\{|x\rangle\}$; generalizations are straightforward.

\subsection{Classical descriptions of M-spaces}

First it needs to be made clear which classical descriptions of M-spaces are considered to be available. It is natural to consider $n$-qubit M-spaces ${\cal M}$ that meet the following requirements:

\begin{itemize}
\item[(i)] ${\cal M}$ has a stabilizer group ${\cal G}=\langle U_1, \dots, U_m\rangle$ with $m=$ poly($n$) generators. A classical description of each $U_i$ is considered to be given as an input; each of these descriptions is assumed to be efficient.
\item[(ii)] Every generator $U_i$ has efficiently computable matrix elements in the following sense. Suppose that $U_i$ acts on the computational basis as \be U_i :\ |x\rangle\ \to \ \lambda_i(x)|\pi_i(x)\rangle\ee for some complex phases $\lambda_i(x)$ and for some permutation $\pi_i$ of the set of $n$-bit strings. Then $U_i$ is said to be efficiently computable if, given any $n$-bit string $x$ as an input, the following two conditions are fulfilled \footnote{See also \cite{Va09} where such operators are called ``efficiently computable basis-preserving''.}:
    \begin{itemize}
    \item There exists a poly$(n)$ time classical algorithm to compute $\pi_i(x)$ and $\pi_i^{-1}(x)$.
    \item There exists a poly$(n, k)$ time classical algorithm to compute the phase $\lambda_i(x)$ up to $k$ bits of precision.
    \end{itemize}
    These conditions entail that, given any row of the matrix $U_i$, it is possible to efficiently compute which matrix element within that row is nonzero and what the value of that matrix element is, and a a similar condition for the columns.
\end{itemize}
The above conditions are met in many cases of interest. In particular we have:

\

{\it Except for the LME states, all M-spaces considered in

section \ref{sect_examples} satisfy (i)-(ii)}.

\

\noindent This statement is easily verified and the arguments are omitted here. LME states generally do not satisfy (i)-(ii) since the phases $\gamma_x$ in (\ref{LME}) may not be efficiently computable. However, all LME states where the function $x\to \gamma_x$ is classically computable in poly$(n, k)$ time up to $k$ bits---which is a large class---do satisfy conditions (i)-(ii).

\subsection{Computational complexity}

Considering $n$-qubit M-spaces described as above one may investigate the computational hardness of a variety of tasks; here an algorithm is considered to be efficient if it runs in poly$(n)$ time. Natural problems are:

\begin{itemize}

\item[(P1)] Decide if the +1 common eigenspace ${\cal M}$ of the generators $U_i$ is nontrivial.

\item[(P2)] Given an M-state $|\psi\rangle$, sample classically from the  distribution $\{|\langle y|\psi\rangle|^2\}$.

\item[(P3)] Given an M-state, compute the expectation value of a $k$-qubit observable for some constant $k$.

\end{itemize}
We will show that one \emph{cannot} hope for efficient classical algorithms for P1-P3 that apply to all M-spaces in general. This is a point where the MSF sharply contrasts with the \emph{Pauli} stabilizer formalism, where many problems of interest can be answered \emph{efficiently} for arbitrary Pauli stabilizer states and codes; this holds in particular problems P1 to P3. Hardness of the  above problems shows that one should look for relevant subclasses of M-spaces for which efficient solutions are possible.

The intractability of P1-P3 in fact holds even for very simple M-spaces viz. those with diagonal, local generators $U_i$:

\vspace{2mm}

{\bf Problem 1}. The input is a set of $n$-qubit diagonal unitary operators $\{U_1, \dots, U_m\}$. Each $U_i$ acts nontrivially on at most 3 qubits and all matrix entries of  $U_i$ (in the computational basis) are either 0, 1 or $-1$. The problem is to decide whether  these operators have a +1 common eigenvector.

\vspace{2mm}

{\bf Problem 2}. The input is a set of diagonal unitary operators $U_i$ as in Problem 1, with the following additional constraint: it is promised that these operators have a unique  (up to a global phase) +1 common eigenstate $|\psi\rangle$, which is thus an M-state. The  problem is to sample classically from the  distribution $\{|\langle x|\psi\rangle|^2\}$.

\vspace{2mm}

{\bf Problem 3}. The input is as in problem 2. The problem is to compute $\langle \psi|Z_i|\psi\rangle$ with accuracy $\epsilon= 1/$poly$(n)$, where $Z_i$ is the Pauli $\sigma_z$ operator acting on qubit $i$.

\vspace{2mm}

\begin{thm}\label{thm_NP_hardness}
None of the problems 1-3 can be solved classically in polynomial time unless P $=$ NP.
\end{thm}

Theorem \ref{thm_NP_hardness} will be proved by reductions to  (variants of) the  satisfiability problem, which is NP-complete. See appendix \ref{app_sect_NPhardness}.

\subsection{Classical simulations}

We focus in more detail on the classical simulation problems P2 and P3. Even though these tasks are intractable in their worst case, efficient solutions exist for subclasses of M-states. Here we provide sufficient criteria for the existence of efficient algorithms.

\begin{thm}\label{thm_class_sim}
Consider an $n$-qubit M-state $|\psi\rangle$. The stabilizer group ${\cal G}$ is described in terms of $m$= poly($n$) efficiently computable generators $\{U_1, \dots, U_m\}$ as above. Let $A$ be a $k$-local observable with $k=O(\log n)$ and $\|A\|\leq 1$. Suppose that the following tasks have efficient classical algorithms:
\begin{itemize}
\item[(a)] Determine any $|x\rangle$ such that $|\psi\rangle\propto |\psi_x\rangle$;
\item[(b)] Generate a uniformly random element in ${\cal O}_x$.
\item[(c)] Given $|y\rangle$, decide if $|y\rangle\in {\cal O}_x$.
\item[(d)] Given $|y\rangle\in {\cal O}_x$, compute $\xi_{x}(y)$.
    \end{itemize}
Then there exists an efficient classical algorithm to sample the  distribution $\Pi:=\{|\langle y|\psi\rangle|^2\}$. Furthermore, then there exists a efficient classical algorithm to estimate $\langle \psi|A|\psi\rangle$ with accuracy $\epsilon=1$/poly$(n)$ with  success probability that is exponentially close to 1.
\end{thm}
{\it Proof:} since $|\psi\rangle$ is a uniform superposition with support ${\cal O}_x$ for some $|x\rangle$ (recall corollary \ref{thm_stab_state}), $\Pi$ is the  \emph{uniform} distribution over this orbit. This shows that efficient classical algorithms for (a) and (b) imply an efficient algorithm to sample $\Pi$.

We consider the second claim. Every $k$-local observable can be written as a linear combination of poly$(n)$ Pauli operators where each coefficient in the linear combination has modulus not greater 1. Therefore it suffices to prove the claim for Pauli operators . For every $a\in \mathbb{Z}_2^n$  consider \be \label{X_and_Z} X(a)&:=& X^{a_1}\otimes \dots \otimes X^{a_n}\nonumber\\Z(a)&:=&Z^{a_1}\otimes\dots\otimes Z^{a_n}.\ee Then every Pauli operator can be written as $\gamma X(a)Z(b)$ for some $a, b$ and some $\gamma\in\{\pm 1, \pm i\}$. We refer to appendix \ref{app_sect_pauli} for some standard properties of Pauli operators. For every $|y\rangle\in {\cal O}_x$ define \be F(y)= \left\{\begin{array}{ll} (-1)^{b^Ty} \xi_x(y)\overline{\xi_x(y+a)}&\mbox{ if } |y+a\rangle\in {\cal O}_x\\ 0 & \mbox{ otherwise.}  \end{array}\right.\ee  Using that $|\psi\rangle\propto |\psi_x\rangle$ and lemma \ref{thm_psi_x} one finds \be\label{chernoff} \langle \psi| X(a)Z(b)|\psi\rangle = \frac{1}{|{\cal O}_x|}\sum_{|y\rangle\in {\cal O}_x} F(y).\ee By a standard Chernoff bound argument (see e.g. the appendix of \cite{Va09}), the sum in (\ref{chernoff}) may be estimated with $1/$poly$(n)$ error with exponentially small failure probability by generating poly$(n)$ random elements $|y_i\rangle\in {\cal O}_x$ and by computing the average of $F(y_i)$. Furthermore owing to assumptions (a)-(d) this procedure can  be implemented in polynomial time. \finpr

Remark that (a) can be approached with theorem \ref{thm_support}. Furthermore, in order to compute $\xi_x(y)$, it suffices to determine \emph{any single} $U\in {\cal G}$ such that $U|x\rangle\propto |y\rangle$, since then $\xi_x(y)$ is simply given by the matrix element $\langle y|U|x\rangle$ (cf. section \ref{sect_basic_xi}). Finally, we point out the following elementary approach to (b):

\begin{lem}\label{thm_sampling}
The following procedure generates a random $|y\rangle\in{\cal O}_x$. First, generate a random permutation $P\in  {\mathcal P}$. Then compute $|y\rangle =P|x\rangle$ and output $y$.
\end{lem}

The proof of lemma \ref{thm_sampling} uses basic group theory arguments and is given in appendix \ref{app_sect_sampling}.  Interestingly, there exists a well-developed theory of (approximately) generating random elements in finite groups and efficient algorithms are available for a variety of groups  (see e.g. \cite{Ba91}). These methods may be thus imported to the study of classical simulations of M-states.

Efficient algorithms for (a)-(d) exist for a variety of M-states. In fact it can be shown that:

\

{\it Except for the LME states, all M-states considered in

section \ref{sect_examples} satisfy (a)-(d) in theorem \ref{thm_class_sim}}.

\

\noindent As before, LME states for which the function $x\to \gamma_x$ can be computed efficiently do satisfy (a)-(d). In section \ref{sect_illustrations} we work out the examples of stabilizer states and quantum double models.

\section{Applications}\label{sect_illustrations}

Throughout this paper we have illustrated our results by means of the simple example of the W states. Here we give some more sophisticated examples viz. the Pauli stabilizer states, the AKLT model and Kitaev's quantum double models.

\subsection{Pauli stabilizer states}

We show how the MSF can be used to rederive some interesting features of Pauli stabilizer states in a new way. In particular:
\begin{itemize}
\item[(i)] We rederive the expansion (\ref{stab_state_uniform_sup}) first proved in \cite{De03}.
\item[(ii)] We show that the conditions (a)-(d) of theorem \ref{thm_class_sim} are fulfilled, thus showing that Pauli stabilizer states can be efficiently  simulated classically in the sense of theorem \ref{thm_class_sim}; this recovers (a variant of) the result \cite{Go98}.
\end{itemize}
Let $|\psi\rangle$ be an $n$-qubit Pauli stabilizer state with Pauli stabilizer group ${\cal G}$.
Recall that every minimal set of generators of ${\cal G}$ contains precisely $n$ Pauli operators \cite{Ni00}. We arbitrarily fix such generators $\{\sigma_1, \dots, \sigma_n\}$. 
Since $Y= iXZ$, every generator can be written as $\sigma_i = \gamma_i X(s^i)Z(t^i)$ where $\gamma_i\in\{ \pm 1, \pm i\}$; recall also the notation (\ref{X_and_Z}). Remark that $\sigma_i$ is a product of a permutation matrix $X(s^i)$ and a diagonal matrix $\gamma_i Z(t^i)$ so that $\bar \sigma_i = X(s^i)$. Let $S$ be the $\mathbb{Z}_2$-linear space generated by the vectors $s^i$. Using (\ref{i},\ref{iii})
one easily finds that \be {\cal P} = \{X(s): s\in S\};\quad {\cal O}_x = \{|x+s\rangle: s\in S\}\ee for every $x\in\mathbb{Z}_2^n$. Invoking corollary \ref{thm_stab_state} these identities immediately imply that $|\psi\rangle$ has the form (\ref{stab_state_uniform_sup}) for some $x\in\mathbb{Z}_2^n$.

We address (ii). First, we show that an $x$ such that $|\psi\rangle\propto |\psi_x\rangle$ can be computed efficiently. Recall the definition of the group ${\cal G}_x$ given in section \ref{sect_support}. Let ${\cal D}$ denote the subgroup consisting of all $\sigma\in {\cal G}$ satisfying $\sigma\propto Z(b)$ for some $b$. It follows from (\ref{i}, \ref{ii}) that ${\cal G}_x={\cal D}$ for every $x$. We now claim:
\begin{lem}\label{thm_pauli_diagonal}
A generating set $\{D_1, \cdots, D_l\}$ of ${\cal D}$ can be determined in poly-time; moreover $D_j = (-1)^{u_j}Z(d^j)$ for some (efficiently computable) $u_j\in \mathbb{Z}_2$ and  $d^j\in\mathbb{Z}_2^n$.
\end{lem}
\noindent This result can be proved using standard Pauli stabilizer arguments; for completeness a proof is given in appendix \ref{app_sect_pauli}. Theorem \ref{thm_support} now implies that $O_x=$ supp$(|\psi\rangle)$ if and only if $D_i|x\rangle = |x\rangle$ for every $i$. Using (\ref{ii}) this is equivalent to requiring that $x^Td^j = u_j$ for every $j$. A solution $x$ to this system of equations can be computed efficiently.

Since we have access to a generating set $\{s^1, \dots, s^n\}$ of $S$, we can efficiently  determine whether $y\in S+x$, given $y$ as input. Also a random element in $x+S$ can easily be generated efficiently.

Finally we show that, given any $y\in x+S$, the phase $\xi_x(y)$ can be computed efficiently. First we compute an arbitrary $a=(a_1, \dots, a_n)\in\mathbb{Z}_2^n$ satisfying $\sum a_i s^i =x+ y$; this regards solving a system of linear equations and can be done efficiently. Using the properties (\ref{i}-\ref{v}) it follows that for such $a$ one has \be \begin{array}{ccc} \sigma_1^{a_1}\dots\sigma_n^{a_n} &\propto& X(\sum a_i s^i)Z(\sum a_i t^i)\\ &&\\&=& X(x+y)Z(\sum a_i t^i).\end{array}\ee It follows that $\sigma_1^{a_1}\dots\sigma_n^{a_n}|x\rangle = \xi |y\rangle$ for some complex phase $\xi$;  by the uniqueness of $\xi_x(y)$ we have $\xi=\xi_x(y) $. Given $a, x$ and $y$ it is straightforward to compute $\xi$ in poly-time.

\subsection{AKLT model with open BCs}

Consider the AKLT model  with open boundary conditions. Let ${\cal M}^{\mbox{\scriptsize{open}}}$ denote the ground level subspace. We will use the MSF to prove the following properties,  first proved in \cite{Af88}:

\begin{itemize}
\item[(i)] The ground level   is 4-fold degenerate;
\item[(ii)] An orthonormal basis of ground states is given by:
\be\label{AKLT_ground_state_open} |\psi_{\mbox{\scriptsize{open}}}^{\sigma}\rangle =\sum \mbox{Tr}\{\sigma\sigma_{a_1}\dots\sigma_{a_n}\} |a_1\dots a_n\rangle.\ee
Here $\sigma\in \{I, X, Y, Z\}$, the sum ranges over all $a_i\in\{0, 1, 2\}$ and $\sigma_0: = X$, $\sigma_1:=Y$ and $\sigma_2:=Z$.  In fact we will prove that (\ref{AKLT_ground_state_open}) is the orbit basis of ${\cal M}^{\mbox{\scriptsize{open}}}$.
\end{itemize}

We prove the results for $n$ even. Let ${\cal G}^{\mbox{\scriptsize{open}}}$ be the group generated by the operators $U_{i,i+1}$. Let $P$ denote the permutation operator obtained by replacing all minus signs in (\ref{AKLT_U}) by plus signs. Then obviously $\bar U_{i,i+1} = P_{i,i+1}$ so that ${\cal P}^{\mbox{\scriptsize{open}}}$ is generated by the operators $P_{i,i+1}$. 
If $n$ is even then it is straightforward to show that ${\cal P}^{\mbox{\scriptsize{open}}}$ has 4 orbits:

\begin{itemize}

\item ${\cal O}_I$ contains all basis states with an even number of $0$s, an even number of 1s
    and an even number of 2s. A representative element is $|a_I\rangle = |0\dots 000\rangle$.

\item ${\cal O}_X$ contains all basis states with an even number of $0$s, an odd number of 1s and an odd number of 2s. A representative element is $|a_X\rangle = |0\dots 012\rangle$.

\item ${\cal O}_Y$ contains all basis states with an even number of $1$s, an odd number of 0s and an odd number of 2s. A representative element is $|a_Y\rangle = |1\dots 102\rangle$.

\item ${\cal O}_Z$ contains all basis states with an even number of $2$s, an odd number of 0s and an odd number of 1s. A representative element is $|a_Z\rangle = |2\dots 201\rangle$.
\end{itemize}
The Pauli matrices satisfy the commutation relations
 \be\label{AKLT_commutation}\begin{array}{lcl} \sigma_0\sigma_1 =  -\sigma_1\sigma_0&\quad &\sigma_1\sigma_2 =  -\sigma_2\sigma_1\\ \sigma_0\sigma_2 =  -\sigma_2\sigma_0 &\quad &\sigma_0\sigma_0 =  \sigma_1\sigma_1 = \sigma_2\sigma_2 (=I).\end{array}\ee
Using these relations it readily follows that each $|\psi_{\mbox{\scriptsize{open}}}^{\sigma}\rangle$ is a +1 eigenstate of all operators $U_{i,i+1}$, so that these 4 states belong to ${\cal M}^{\mbox{\scriptsize{open}}}$. Using the commutation relations (\ref{AKLT_commutation}) and the fact that $X$, $Y$ and $Z$ are traceless operators, one also finds that $|\psi_{\mbox{\scriptsize{open}}}^{\sigma}\rangle$ has support ${\cal O}_{\sigma}$. This shows that these four states are orthogonal. Therefore the dimension of ${\cal M}^{\mbox{\scriptsize{open}}}$ is at least 4 and  each orbit ${\cal O}_{\sigma}$ belongs to the support of ${\cal M}^{\mbox{\scriptsize{open}}}$. Owing to corollary \ref{thm_dimension} the dimension is at most the total number of orbits of ${\cal P}^{\mbox{\scriptsize{open}}}$, which equals 4.  This shows that ${\cal M}^{\mbox{\scriptsize{open}}}$ is 4-dimensional with basis (\ref{AKLT_ground_state_open}).

Finally we show that $|\psi_{\mbox{\scriptsize{open}}}^{\sigma}\rangle$ is the orbit state determined by $|a_{\sigma}\rangle$. Since all four orbits ${\cal O}_{\sigma}$ belong to the support of ${\cal M}^{\mbox{\scriptsize{open}}}$, the orbit basis consists of the four corresponding orbit states. As $|\psi_{\mbox{\scriptsize{open}}}^{\sigma}\rangle$ belongs to ${\cal M}^{\mbox{\scriptsize{open}}}$, this state must be a linear combination of the states in the orbit basis. Since $|\psi_{\mbox{\scriptsize{open}}}^{\sigma}\rangle$ has support ${\cal O}_{\sigma}$ as argued above, the latter is only possible if $|\psi_{\mbox{\scriptsize{open}}}^{\sigma}\rangle$  \emph{is}  the orbit state determined by $|a_{\sigma}\rangle$.

By using the commutation relations (\ref{AKLT_commutation}) and by applying the definition of the phases $\xi_x(y)$ it can also be shown directly that $|\psi_{\mbox{\scriptsize{open}}}^{\sigma}\rangle$  is the orbit state determined by $|a_{\sigma}\rangle$. This argument is omitted here.

\subsection{AKLT model with periodic BCs}

Consider the AKLT model with periodic boundary conditions and $n$ even.
Let ${\cal M}^{\mbox{\scriptsize{per}}}$ denote the ground level subspace. We rederive the following properties,  first proved in \cite{Af88}:
\begin{itemize}
\item[(i)] The ground level is non-degenerate;
\item[(ii)] The unique ground state is (using an analogous notation as in (\ref{AKLT_ground_state_open}))
\be\label{AKLT_ground_state_periodic} |\psi_{\mbox{\scriptsize{per}}}\rangle \propto\sum \mbox{Tr}\{\sigma_{a_1}\dots\sigma_{a_n}\} |a_1\dots a_n\rangle.\ee
\end{itemize}

Using the commutation relations (\ref{AKLT_commutation}) it is straightforward to show that $|\psi_{\mbox{\scriptsize{per}}}\rangle$ belongs to ${\cal M}^{\mbox{\scriptsize{per}}}$. As above, ${\cal P}^{\mbox{\scriptsize{per}}}$ is generated by the permutation matrices $P_{i ,i+1}$ (now imposing periodic boundary conditions). Furthermore one may verify that ${\cal P}^{\mbox{\scriptsize{per}}}$ has the same four orbits as in the open boundary conditions case. Note also that $|\psi_{\mbox{\scriptsize{per}}}\rangle$ has support ${\cal O}_I$ so that ${\cal O}_I\subseteq$ supp$({\cal M}^{\mbox{\scriptsize{per}}})$. We show that none of the other orbits are contained in the support. To do so define the operators \be A &=& U_{n, n-1} U_{n-1, n-2} \dots U_{2,1} U_{1,n} \nonumber\\ B &=& U_{n-2, n-1} A.\ee One can then verify that $B|a_{\sigma}\rangle = -|a_{\sigma} \rangle$ for every $\sigma= X, Y, Z$.
Since $B\in{\cal G}^{\mbox{\scriptsize{per}}}$, owing to theorem \ref{thm_support} this implies that none of the orbits ${\cal O}_X, {\cal O}_Y$ or ${\cal O}_Z$ are contained in the support of ${\cal M}^{\mbox{\scriptsize{per}}}$. It follows that the support coincides with the single orbit ${\cal O}_I$. Invoking corollary \ref{thm_dimension} then implies that ${\cal M}^{\mbox{\scriptsize{per}}}$ is one-dimensional.

\subsection{Quantum double models}

Consider a quantum double model defined on a sphere (cf. appendix \ref{app_sect_qd}). Such a system has a unique ground state $|\psi_{\mbox{\scriptsize{qd}}}\rangle$ \cite{Ki03}, which is thus an M-state. Letting ${\cal G}$ be the stabilizer group of this state, here we show: \begin{itemize}
\item ${\cal G}$ meets requirements (a)-(d) of theorem \ref{thm_class_sim}, showing that $|\psi_{\mbox{\scriptsize{qd}}}\rangle$ can be simulated classically in the sense of theorem \ref{thm_class_sim}.
\end{itemize}
The reference work \cite{Ki03} can also be used to show that standard basis measurements can be simulated classically (although it is not explicitly discussed there). Efficient simulations of local expectation values were previously achieved using tensor network methods \cite{Ag08, Vi08}.

First we describe the stabilizer group. Recall the definition of the operators $U_p$ and $V_v(k)$ defined in appendix \ref{app_sect_qd}.    Since $U_p$ is diagonal and $V_v(k)$ is a permutation matrix, the group ${\cal G}$ generated by these operators is pure.   Note that the $U_p$ mutually commute since these are diagonal operators. Furthermore it is easily  verified that $[V_v(k), U_p]=0$ and $[V_v(k), V_w(l)]=0$ for all vertices $v\neq w$, for every plaquette $p$ and for every $k, l\in G$. It follows that a general element $U\in {\cal G}$ has the form $U = PD$ where \be\label{quantum_double_PD} P&=& \prod_v V_v(k_v)\quad \mbox{ where } k_v\in G\nonumber \\ D&=& \prod_p U_p^{x_p}\quad \mbox{ where } x_p\in\{0, 1\}.\ee Since $P$ is a permutation matrix and $D$ is diagonal, one has $\bar U=P$.

Next we determine the support of $|\psi_{\mbox{\scriptsize{qd}}}\rangle$. A standard basis state has the form $|\mathbf{g}\rangle=\bigotimes|g_e\rangle$ where the product is over all edges $e$ and where $g_e\in G$. Let ${\cal S}$ be the set of all $|\mathbf{g}\rangle$ satisfying  $U_p|\mathbf{g}\rangle=|\mathbf{g}\rangle$ for all plaquettes $p$. We claim that supp$(|\psi_{\mbox{\scriptsize{qd}}}\rangle) = {\cal S}$. To see this, first note that theorem \ref{thm_support} shows that supp$(|\psi_{\mbox{\scriptsize{qd}}}\rangle) \subseteq {\cal S}$. To prove the reverse inclusion, consider an arbitrary $U=PD\in {\cal G}$ as in (\ref{quantum_double_PD}).  If $|\mathbf{g}\rangle\in{\cal S}$ then $D|\mathbf{g}\rangle=|\mathbf{g}\rangle$. Consequently, \be \langle \mathbf{g}|U|\mathbf{g} \rangle = \langle \mathbf{g}|PD|\mathbf{g} \rangle = \langle \mathbf{g}|P|\mathbf{g}\rangle \in\{0, 1\},\ee where the last inclusion holds since $P$ is a permutation matrix. Invoking theorem \ref{thm_support}  proves the claim.

Recall that $|\psi_{\mbox{\scriptsize{qd}}}\rangle$ is an M-state and that ${\cal G}$ is pure. Corollary \ref{thm_pure} implies that $|\psi_{\mbox{\scriptsize{qd}}}\rangle$ is the equal superposition over ${\cal S}$, and that ${\cal S}$ must be an orbit of ${\cal P}$. Since one manifestly has $|\mathbf{e}\rangle\in{\cal S}$ (where $\mathbf{e}$ has the neutral element $e$ in all its entries), it follows that ${\cal S}={\cal O}_{\mathbf{e}}$. Consequently,  $|\psi_{\mbox{\scriptsize{qd}}}\rangle$ coincides with the orbit state $|\psi_{\mathbf{e}}\rangle$. This shows that (a) in theorem \ref{thm_class_sim} is fulfilled.

To show (b) we use lemma \ref{thm_sampling}. Since an arbitrary $P\in{\cal P}$ has the form given in (\ref{quantum_double_PD}), a random $P$ can efficiently be generated by generating a random $k_v\in G$ for every vertex $v$. Applying $P$ to $|\mathbf{e}\rangle$ yields a random state in ${\cal O}_{\mathbf{e}}$.

Condition (c) holds since the support ${\cal O}_{\mathbf{e}}={\cal S}$ is defined in terms of polynomially many  constraints $U_p|\mathbf{g}\rangle = |\mathbf{g}\rangle$, each of which is easily verified. Finally, since $|\psi_{\mbox{\scriptsize{qd}}}\rangle$ is an equal superposition state, one has $\xi_{\mathbf{e}}(\mathbf{g})=1$ for every $\mathbf{g}\in{\cal O}_{\mathbf{e}}$.

\begin{acknowledgements}
I am very grateful to Miguel Aguado, Oliver Buerschaper, Wolfgang D\"ur, Richard Jozsa, Barbara Kraus, Robert Raussendorf and Hong-Hao Tu for very helpful discussions, and to Wolfgang D\"ur, Richard Jozsa and Barbara Kraus for very useful comments on the manuscript.
\end{acknowledgements}
\appendix

\section{Examples of M-states and -spaces}\label{app_sect_further_examples}

\subsection{Pauli stabilizer formalism for qudits}

Generalizations of Pauli $X$ and $Z$ matrices exist for $d$-level systems (``qudits'') with arbitrary $d$ as follows:
\be X_d = \sum_{x=0}^{d-1} |x+1\rangle\langle x|; \quad Z_d = \sum_{x=0}^{d-1} e^{2\pi i x/d}|x\rangle\langle x|.\ee
Here $\{|0\rangle, \dots|d-1\rangle\}$ is the standard basis and $x+1$ is computed modulo $d$. Based on these definitions one may consider generalized Pauli operators acting on $n$ qudits and the associated stabilizer states and codes. It is straightforward that generalized Pauli operators are also monomial unitary operators relative to the standard basis, so that stabilizer states and codes for qudits are M-states and-spaces, respectively.

\subsection{Quantum double models}\label{app_sect_qd}

Consider a finite group $G$ and a Hilbert space ${\cal H}_{\mbox{\scriptsize{loc}}}$ with basis $\{|g\rangle: g\in G\}$ where basis vectors are labeled by group elements. Furthermore consider a 2D square lattice $\Lambda$ arranged on a sphere, where with each edge a Hilbert space ${\cal H}_{\mbox{\scriptsize{loc}}}$ is associated (generalizations to other 2D lattices are possible). Finally we assign an arbitrary orientation to each edge of the lattice. Now consider the Hamiltonian $ H_{\mbox{\scriptsize{qd}}} = -\sum B_p - \sum A_v$ where the first (second) sum runs over all plaquettes $p$ (vertices $v$) of the lattice. The operators
$B_p$ and $A_v$ are mutually commuting projectors with support along the boundary of the plaquette $p$ and on the edges adjacent
to $v$, respectively. To define $B_p$, consider the  edges $(e_1, e_2, e_3, e_4)$ of $p$ when traversing this plaquette counter-clockwise starting from some arbitrary vertex.  The  signature $s_i$ of the edge $e_i$ is $+1$ if  $e_i$ is pointing along the counter-clockwise sense, and $-1$ if it is pointing in the opposite sense. Now consider a basis state $|g_1, g_2, g_3, g_4\rangle$ where $g_i\in G$ is associated with the system on edge $e_i$. Then $B_p$ acts as follows:
\be B_p|g_1, g_2, g_3, g_4\rangle = \delta(g_4^{s_4} g_3^{s_3} g_2^{s_2} g_1^{s_1}) |g_1, g_2, g_3,g_4\rangle,\ee where $\delta(e)=1$ and $\delta(k)=0$ for every $e\neq k\in G$ (where $e$ is the neutral element of $G$).
To define $A_v$ consider the four edges $(f_1, f_2, f_3, f_4)$ incident on $v$. The signature $t_i$ of $f_i$ is now $+1$ if $f_i$ is pointing towards $v$ and $-1$ if it is pointing away from $v$.  Consider a basis state $|g_1, g_2, g_3, g_4\rangle$ where $g_i\in G$ is associated with the system on edge $f_i$. If $k\in G$ define \be k*g_i=\left\{\begin{array}{cl} kg_i & \mbox{ if }  t_i = 1\\ g_i k^{-1} & \mbox{ if }  t_i = -1 \end{array}\right. \ee Now define \be A_v|g_1, g_2, g_3, g_4\rangle = \frac{1}{|G|}\sum_{k\in G} |k*g_1, k*g_2, k*g_3, k*g_4\rangle.\nonumber\\ \ee
The ground state subspace ${\cal M}_{\mbox{\scriptsize{qd}}}$ of $H_{\mbox{\scriptsize{qd}}}$ consists of those  $|\psi\rangle$ which are joint +1 eigenvectors of all projectors $B_p$ and $A_v$. To show that ${\cal M}_{\mbox{\scriptsize{qd}}}$ is an M-space, define \be U_p &=& 2B_p - I\nonumber \\ V_v(k)|g_1, g_2, g_3, g_4\rangle &=& |k*g_1, k*g_2, k*g_3, k*g_4\rangle,\nonumber\\\ee where $k\in G$. Then $U_p$ is a diagonal unitary operator and $V_v(k)$ is a permutation operator. It is immediate that the $+1$ eigenspaces of $U_p$ and $B_p$ coincide. Furthermore the $+1$ eigenspace of $A_v$ coincides with the +1 joint eigenspace of the set ${\cal A}_v:=\{V_v(k): k\in G\}$. To see this, note that (a) ${\cal A}_v$ is a group and (b) $A_v$ is the averaging operator over this group.  It follows that (cf. appendix \ref{app_sect_rho}) $A_v$ is the projector onto the +1 common eigenspace of ${\cal A}_v$. In other words  the $+1$ eigenspace of $A_v$ coincides with the +1 joint eigenspace of ${\cal A}_v$.

In conclusion, ${\cal M}_{\mbox{\scriptsize{qd}}}$ coincides with the +1 common eigenspace of the operators $U_p$ and $V_v(k)$ where $p$ ranges over all plaquettes, $v$ ranges over all vertices and $k$ over the entire group $G$. This shows that ${\cal M}_{\mbox{\scriptsize{qd}}}$ is an M-space.

\subsection{Laughlin wavefunction at $\nu=1$}

The Laughlin wave function for $n$ particles and filling fraction $\nu=1$ is \be |\psi_{\mbox{\scriptsize{l}}}\rangle = \sum_{a_1, \dots, a_n =0}^{n-1} \epsilon^{a_1\dots a_n} |a_1\rangle\dots|a_n\rangle\ee Here $|a_i\rangle$ is a wavefunction for particle $i$ given by \be \langle z|a_i\rangle =\frac{1}{\sqrt{\pi a!}} z^{a_i} e^{-|z|^{2}/2}\ee where $z$ is a complex number encoding the position of the particle in a two-dimensional plane. Furthermore $\epsilon$ is the Levi-Civita completely antisymmetric tensor in $n$ dimensions. Let $S_{ij}$ be the operator which swaps systems $i$ and $j$.
It is straightforward that each operator $-S_{ij}$ is unitary and monomial and that $|\psi_{\mbox{\scriptsize{l}}}\rangle$ is the unique +1 common eigenvector of these operators. Thus $|\psi_{\mbox{\scriptsize{l}}}\rangle$ is an M-state.

\subsection{Coherent probabilistic computations}
Consider a poly-size classical circuit composed of reversible gates (e.g. Toffoli gates and NOT gates). The circuit acts on an $n$-bit input string where, say, the first $k$ bits are uniformly random and the last $n-k$ input bits are initialized in 0. Let $\{\pi_x: x\in\mathbb{Z}_2^n\}$ denote the output probability distribution over the set of $n$-bit strings and define the state $|\pi\rangle = \sum \sqrt{\pi_x}|x\rangle.$ Let $\tilde C$ denote the natural translation of $C$ into an $n$-qubit quantum circuit. Furthermore let $X_i$ and $Z_j$ denote the Pauli $X$ and $Z$ operators acting on qubit $i$ and $j$, resp., and define $P_i = \tilde CX_i{\tilde C}^{\dagger}$ and $D_j={\tilde C}Z_j{\tilde C}^{\dagger}$. These operators are unitary and monomial (relative to the computational basis) and have $|\pi\rangle$ as unique +1 common eigenstate. Thus $|\pi\rangle$ is an M-state.

\subsection{Coset states of finite Abelian groups}

Consider a Hilbert space with basis $\{|g\rangle :g\in G\}$ where basis vectors are labelled by the elements of a finite abelian group $G$. Let $\chi_g$ be the character of $G$ canonically associated to $g$ and define the operators \be X(g) :\ |x\rangle\to |x+g\rangle;\quad Z(g) :\ |x\rangle\to \chi_g(x)|x\rangle,\ee for every $g, x\in G$. Remark that these operators are unitary and monomial.
Consider a subgroup $H$ of $G$ with generating set $\{h^1, \dots, h^n\}$. The dual group $H^{\perp}$ is the set of all $k$ satisfying $\chi_h(k)=1$ for every $h\in H$; consider a generating set $\{k^1, \dots, k^m\}$. Now define the ``coset state'' $|H+x\rangle =\sum |x+h\rangle$ where the sum is over all $h\in H$. This state is the unique +1 common eigenvector of the operators $X(h^i)$ and $Z(k^j)$ where $i$ ranges from 1 to $n$ and where $j$ ranges from 1 to $m$. Thus such coset states are M-states.  Remark that these states play an important role in the context of quantum algorithms viz. the Abelian hidden subgroup problem.

\subsection{W-states and Dicke states}

Consider a system of $n$ qubits with computational basis ${\cal B}$. Fix $k\in\{1, \dots, n\}$ and define  $|D^k\rangle$ to be the (properly normalized) equal superposition over all basis states $|x\rangle$ for which the bit string $x$ has hamming weight $k$. Note that $|D^1\rangle=|W\rangle$ is the W state. Denote $T_k = \bar\alpha^k\cdot \Lambda^{\otimes n}$, where $\Lambda = $ diag$(1, \alpha)$ and $\alpha = e^{\frac{2\pi i}{n}}$.
Further, let $S_i$ denote the  SWAP gate acting on qubits $i$ and $i+1$. The operators $T_k$ and $S_i$ are unitary and monomial and have $|D^k\rangle$ as unique +1 common eigenstate. Thus $|D^k\rangle$ is an M-state.

\section{The projector $\rho$}\label{app_sect_rho}

Let $\pi = |{\cal G}|^{-1}\sum U$ be the averaging operator over ${\cal G}$. We prove that $\pi = \rho$. Since $\pi$ is defined as a sum over all elements in ${\cal G}$, one has $U\pi = \pi$ for every $U\in{\cal G}$. Consequently, \be \pi^2 = \frac{1}{|{\cal G}|} \sum_{U\in{\cal G}} U \pi = \pi.\ee Moreover, $\pi = \pi^{\dagger}$ since for every $U\in{\cal G}$ one has $U^{\dagger}\in{\cal G}$ as ${\cal G}$ is a group. Thus $\pi^2 = \pi= \pi^{\dagger}$ i.e. $\pi$ is an orthogonal projector. Now let ${\cal W}$  denote the $+1$ eigenspace of $\pi$. First note that the definition of $\pi$ immediately implies that $\pi|\psi\rangle = |\psi\rangle$ for every $|\psi\rangle\in{\cal M}$, showing that ${\cal M}\subseteq{\cal W}$. Furthermore, consider an arbitrary $|\varphi\rangle\in{\cal W}$ i.e. $\pi|\varphi\rangle = |\varphi\rangle$. Since $U\pi = \pi$ for every $U\in{\cal G}$, it follows that \be U|\varphi\rangle = U\pi|\varphi\rangle = \pi|\varphi\rangle = |\varphi\rangle,\ee showing that $|\varphi\rangle\in{\cal M}$. This proves that ${\cal W}\subseteq {\cal M}$ and we conclude that ${\cal W} = {\cal M}$. Thus $\rho$ and $\pi$ are orthogonal projectors on the same space, so that these operators are equal.

To prove (\ref{U_rho=rho}), consider an arbitrary fixed $U\in {\cal G}$. Since ${\cal G}$ is a group, the sum $\sum_{V\in G} UV$ is precisely the sum over all elements of ${\cal G}$. Using (\ref{rho}) it follows  that $U\rho=\rho$.

\section{Proof of theorem \ref{thm_NP_hardness}}\label{app_sect_NPhardness}

Problem 1 is shown to be NP-hard by relating it to 3-{\sc satisfiability} (3SAT). In the latter problem, the input is a Boolean function $f(x)$ (where $x=x_1\cdots x_n$ is a bit string) written in conjunctive normal form with 3 variables per clause (3-CNF). That is $f(x) = f_1(x)\wedge \cdots \and \wedge f_k(x)$ where each clause $f_i$ is a disjunction of three literals; a literal is either a variable or its negation. An example of a clause is $x_3\vee \neg x_6 \vee x_7$. The 3SAT problem is to decide whether $f$ is satisfiable i.e. if there exists an $x$ such that $f(x)=1$.   To reduce 3SAT to Problem 1, consider an $n$-qubit system and for each $i$ define the unitary operator $U_{i}$ which maps $|x\rangle$ to itself if $f_i(x)=1$ and to $-|x\rangle$ otherwise. It is then easy to verify that the operators $U_1, \dots, U_k$ have a +1 common eigenvector if and only if $f$ is satisfiable. Moreover every $U_i$ is unitary and diagonal an acts on at most 3 qubits.

As for Problems 2 and 3 we consider the following variant of 3SAT. The input is again a Boolean function $f$ in 3-CNF form; in addition it is promised that there exists a unique $x^*$ such that $f(x^*)=1$. The problem is to determine $x^*$. This problem is known to be NP-hard (under randomized reductions) \cite{Va85}. To reduce it to problem 2, consider the translations of the clauses $f_i$ into diagonal unitary operator $U_i$ as above. Owing to the promise on $f$, it follows that the state $|\psi\rangle:=|x^*\rangle$ is an M-state with stabilizer group $\langle U_1, \cdots, U_k\rangle$. Sampling the distribution $\Pi$ is equivalent to determining $x^*$.

As for Problem 3, note that estimating the $n$ expectation values $\langle\psi|Z_i|\psi\rangle$ in polynomial time with the prescribed accuracy allows one to determine $x^*$ in polynomial time as well.

\section{Proof of lemma \ref{thm_sampling}}\label{app_sect_sampling}

Clearly, the procedure in the lemma can only output elements lying within ${\cal O}_x$. Note that every $|y\rangle$ in this orbit occurs with probability \be \mbox{Prob}(|y\rangle) = |\{P\in  {\mathcal P}:|y\rangle =P|x\rangle\}/|{\cal G}|.\ee Let ${\mathcal Q}$ consist of all $P\in\ {\cal P}$ satisfying $P|x\rangle =|x\rangle$. Then ${\cal Q}$ is easily shown to be a subgroup of ${\cal G}$. Further, let $P_y\in{\cal G}$ be a permutation satisfying $ P_y|x\rangle = |y\rangle$ and consider the set $P_y{\mathcal Q}$ consisting of all products $P_yP$ with $P\in {\mathcal Q}$ (formally, this is the left coset of ${\mathcal Q}$ defined by $P_y$). We now claim that \be\label{sets_equality} P_y{\mathcal Q}= \{P\in {\cal P}: P|x\rangle=|y\rangle\}.\ee The inclusion $\subseteq$ holds trivially. To prove the converse inclusion, note that every $P$ satisfying $P|x\rangle=|y\rangle$ can be written as $P=P_y[P_y^{-1} P]$, where $P_y^{-1} P\in {\mathcal Q}$. This shows (\ref{sets_equality}). Therefore $\mbox{Prob}(|y\rangle)$ is equal to $|P_y{\mathcal Q}|/|{\mathcal P}|$. Note however that by definition the set $P_y{\mathcal Q}$ has the same cardinality as ${\mathcal Q}$ for every $y$. Thus $\mbox{Prob}(|y\rangle)$ is independent of $y$ so that this probability distribution is uniform, as desired.

\section{Pauli operators}\label{app_sect_pauli}

For every  $a, a', x\in \mathbb{Z}_2^n$, the following well-known identities are easily verified: \be X(a)|x\rangle&=& |x+a\rangle \label{i}\\
Z(a)|x\rangle&=& (-1)^{x^Ta} |x\rangle \label{ii}\\
X(a)X(a') &=& X(a+a') \label{iii}\\Z(a)Z(a')&=& Z(a+a')\label{iv}\\
X(a)Z(a')&=& (-1)^{a^Ta'} Z(a')X(a)\label{v}\ee
If $\sigma= i^{k}X(s)Z(t)$ (where $k\in\{0, 1, 2, 3\}$ and $s, t\in\mathbb{Z}_2^n$) is a Pauli operator we will call $(k, s, t)$ the label of $\sigma$.

{\it Proof of lemma \ref{thm_pauli_diagonal}:}  we will need take the following technicality in account. Remark that $\sigma^2 = \alpha I$ for every Pauli operator $\sigma$, where $\gamma=\pm 1$ . Now suppose that there exists $\sigma\in {\cal G}$ where $\gamma$ is \emph{not} equal to 1. Then, since $\gamma I =\sigma^2\in {\cal G}$ one would have $\gamma I|\psi\rangle = |\psi\rangle$ so that $|\psi\rangle = 0$, leading a contradiction. Thus we conclude that $\sigma^2=I$ for every $\sigma\in {\cal G}$.

Since the $\sigma_i$ mutually commute and square to the identity, every operator in ${\cal G}$ can be parameterized as $\sigma(a):= \sigma_1^{a_1}\dots \sigma_n^{a_n}$, where $a=(a_1, \dots, a_n)\in\mathbb{Z}_2^n.$
Using (\ref{i}-\ref{v}) it follows that \be\label{prop2_U_x} \begin{array}{c}\sigma(a) \propto X(\sum a_i s^i)Z(\sum a_i t^i)\end{array}\ee This implies that $\sigma(a)\in {\cal D}$ if and only if $a$ solves the linear equation $\sum a_i s^i=0$. Let $\{a^1, \dots, a^k\}$ denote a basis of solutions; such a basis can be computed efficiently. Thus $\sigma(a)\in{\cal D}$ if and only if $a = \sum y_j a^j$ for some $y_j\in\mathbb{Z}_2$. Furthermore, using the definition of $\sigma(a)$ it is straightforward to show that \be \begin{array}{c}\sigma( \sum y_j a^j) = \sigma(a^1)^{y_1}\dots \sigma(a^k)^{y_k}.\end{array}\ee This shows that the operators $D_j:=\sigma(a^j)$ form a generating set of ${\cal D}$. Since $\sigma(a^j)$ is a product of at most $n$ Pauli products, the label of each $\sigma(a^j)$ can be computed in poly$(n)$ time. Finally, since every element of ${\cal G}$ squares to the identity it follows that $D_j = (-1)^{u_j}Z(d^j)$ for some $u_j\in \mathbb{Z}_2$  and some $d^j\in\mathbb{Z}_2^n$.\finpr

\end{document}